\documentclass[10pt]{revtex4}
\usepackage{amssymb,epsf}
\usepackage{latexsym}

\begin{document}

\title{From the Komar mass and entropic force scenarios to the Einstein field equations on the AdS brane}
\author{H. Moradpour$^{1}$\footnote{h.moradpour@riaam.ac.ir} and A. Sheykhi$^{1,2}$\footnote{asheykhi@shirazu.ac.ir}}
\address{$^1$ Research Institute for Astronomy and Astrophysics of Maragha
         (RIAAM), P.O. Box 55134-441, Maragha, Iran\\
         $^2$ Physics Department and Biruni Observatory, College of
Sciences, Shiraz University, Shiraz 71454, Iran}

\begin{abstract}
By bearing the Komar's definition for the mass, together with the
entropic origin of gravity in mind, we find the Einstein field
equations in $(n+1$)-dimensional spacetime. Then, by reflecting the
($4+1$)-dimensional Einstein equations on the ($3+1$)-hypersurface,
we get the Einstein equations onto the $3$-brane. The corresponding
energy conditions are also addressed. Since the higher dimensional
considerations modify the Einstein field equations in the
($3+1$)-dimensions and thus the energy-momentum tensor, we get a
relation for the Komar mass on the brane. In addition, the
strongness of this relation compared with existing definition for
the Komar mass on the brane is addressed.
\end{abstract}
\maketitle

\section{Introduction}
The discovery of the thermodynamic roots of gravity comes back to
the works of Bekenstein and Hawking \cite{B1,H1,H2}. These
attempts were followed \cite{B2,D1,u1,s1} and finally formed the
backbone of the well-known paper by Jacobson \cite{Jac}. Indeed,
Jacobson has considered the same relation between the horizon
entropy and its surface area as the Bekenstein bound \cite{B1},
and showed that for the static spacetimes the Einstein field
equations on the horizon is equal to the thermodynamic identity
$\delta Q=T \delta S$. This setup were also extended to $f(R)$
gravity \cite{Elin}. In fact, by using the first law of
thermodynamics on the horizon, it was proved that the
gravitational field equations can be rebuilt in a wide range of
theories \cite{Pad}. The same deductions for cosmological setups
\cite{Cai2,Cai3,CaiKim,W1,W2,Cai33,Shey0,Shey01} and braneworld
scenarios \cite{Shey02,Shey1,C1,Shey2,Shey22,Shey23} are valid. A
comprehensive review can be found in Ref. \cite{Pad0}.

Thermodynamic aspects of the gravity can help us to have a better
understanding of the nature of spacetime and the gravity.
Padmanabhan claimed that spacetime includes unknown structure in
the microscopic scales inducing the degrees of freedom for the
spacetime, and their statistical description yields the gravity
\cite{Padm}. Another parallel approach suggested by Verlinde
\cite{Ver}. He showed that the tendency of systems to increase
their entropy may lead to the emergence of the gravity. Therefore,
gravity is not a fundamental force and can be considered as a
secondary effect. This approach is called entropic force and has
attracted a lot of investigations
\cite{Cai4,Cai41,Smolin,Li,Tian,Myung1,Vancea,Modesto,
Sheykhi1,BLi,Sheykhi2,Sheykhi21,Sheykhi22,Sheykhi23,Ling,Sheykhi24,Gu,Miao1,other,mann,SMR}.

String theory, as a promising approach to quantum gravity, predicts
eleven dimensions for the spacetime \cite{Pol}. The AdS/CFT
correspondence conjecture relates an $n$-dimensional conformal field
theory to an ($n+1$)-dimensional gravity theory in an anti-de Sitter
(AdS) space \cite{Mad,Mad1}. Indeed, such generalization to the
$n$-dimensions is due to the holographic principle \cite{Sus,Bous}.
Based on these motivations, bearing the Komar's $n$-dimensional
definition of mass \cite{Cai4,Cai41} in mind and using the entropic
force approach, the Friedmann equations of the ($n+1$)-dimensional
gravity theories including the Einstein, Gauss-Bonnet and Lovelock
theories are obtainable \cite{SMR}.

Another striking motivation for studying the higher dimensional
gravity comes from braneworld scenarios. In these scenarios there is
a ($4+1$)-dimensional spacetime and unlike other fields, gravity can
propagate in the fifth dimension. It was argued that the braneworld
scenarios can explain the weakness of the gravity compared with the
other fundamental forces \cite{RS1,RS}. By taking the entropic
origin of the gravity into account, authors of \cite{Ling} tried to
find the Friedmann equations on the brane. In addition, author of
\cite{Sheykhi24} has considered corrected entropy of the horizon and
found the modified Friedmann equations on the brane from the
entropic force. In these approaches \cite{Ling,Sheykhi24}, authors
have introduced a same definition for the Komar mass on the brane.
In this paper, based on the Komar definition for the mass in
$n$-dimensions, we try to build the ($n+1$)-dimensional Einstein
field equations using the entropic force approach. In continue we
point to this fact that by following the covariant approach of the
authors of \cite{shim}, one can reach to the Einstein field
equations onto the brane. We also study the validity of some energy
conditions on the brane. We also get a relation for the Komar mass
on the brane which is in accordance with the higher dimensional
modifications to the Einstein equations onto the brane. In addition,
we compare our result for the Komar mass on the brane with the
previous definition introduced in \cite{Ling,Sheykhi24}. Finally, we
point to weaknesses of the Einstein field equations derived by using
the entropic force scenario together with the Komar mass definition
used in \cite{Ling,Sheykhi24}.

\section{Gravity in ($n+1$)-dimensions}
According to the Verlinde's proposal, the tendency of the systems
to increase their entropy leads to the emergence of gravity
\cite{Ver}
\begin{equation}\label{F}
F=T\frac{\triangle S}{\triangle x}.
\end{equation}
In this approach there is a surface sphere (holographic screen) with
radius $r$ which encloses the source of energy. The holographic
principle implies $S\sim A$, where $A$ is the surface area of the
holographic screen \cite{mann}. Since the gravitational information
of the energy source is distributed over $N$ bits on the holographic
screen, we also have $S\sim N$ \cite{mann}. In addition, the entropy
of the gravitational system, according to the Bekenstein argument
\cite{B1}, is given by
\begin{eqnarray}\label{entropy1}
S=\frac{A}{ 4 \ell_p^2},
\end{eqnarray}
where $\ell_p$ is the Plank's length and $A$ is the surface area
of the three-dimensional holographic screen \cite{Ver}. Since the
possible maximum number of the bits on the three dimensional
holographic screen is given by
\begin{eqnarray}\label{n}
N=\frac{A}{\ell_p^2},
\end{eqnarray}
we reach $S=\frac{N}{4}$ for the relation between the entropy and
the number of the bits on the three dimensional holographic surface
\cite{Ver,mann}. The Unruh temperature associated with the
holographic screen can be written \cite{u1},
\begin{eqnarray}
k_B T=\frac{\hbar}{2\pi c}a,
\end{eqnarray}
where $a$ is the acceleration of the Unruh observer. One can
generalize this temperature to every accelerated observer when $F
\neq 0$ \cite{Ver}. In Eq. (\ref{F}), $\Delta x$ is the
displacement of the test mass ($m$) from the holographic screen.
When $\Delta x$, is of order of Compton wavelength
$\lambda_m=\frac{\hbar}{mc}$, the test mass will be absorbed by
the source \cite{Ver}. In order to generalize our study to the
arbitrary dimensions, we consider an ($n+1$)-dimensional spacetime
with timelike Killing vector $\xi^{\alpha}$, where its metric is
$g_{\mu \nu}$. So, $\phi=-\frac{1}{2}\log\xi^{\alpha}\xi_{\alpha}$
is the generalization of the Newton's potential \cite{Ver}. The
mass $M$ induces a holographic screen $\Sigma_n$ at distance $r$
and for the volume and the area of this $n$-sphere, we have
\begin{equation}
V_n=\Omega_n R^n,  \  \   \   \Sigma_n=n\Omega_{n}R^{n-1},
\end{equation}
where
\begin{equation}
\Omega_{n}=\frac{\pi^{n/2}}{\Gamma(\frac{n}{2}+1)},  \  \  \   \
\Gamma(\frac{n}{2})=(\frac{n}{2}-1)!.
\end{equation}
The $(n+1)$-dimensional gravitation constant may be written as
\cite{mann,SMR}
\begin{equation}\label{G}
G_{n+1}=2 \pi^{1-n/2}\Gamma\left(\frac{n}{2}\right)\frac{c^3\ell
_p^{n-1} }{\hbar}.
\end{equation}
Generalization of (\ref{n}) to $n$-dimension yields \cite{mann,ppp}
\begin{equation}\label{N}
\Sigma_n=N\ell_p^{n-1},
\end{equation}
while we still have
\begin{eqnarray}\label{entropy}
S\sim \Sigma_n,
\end{eqnarray}
which is due to the holographic principle \cite{ppp,mann}. Consider
the total energy of the system as
\begin{equation}\label{Ec}
E=M c^2,
\end{equation}
which is regularly distributed over the $N$ bits. According to the
equipartition law of energy we have \cite{Pad1}
\begin{equation}\label{E}
E=\frac{1}{2}Nk_B T.
\end{equation}
By combining this equation with (\ref{Ec}) and employing the
general equipartition law of the energy, we arrive at \cite{Ver}
\begin{equation}\label{mass}
M=\frac{1}{2c^2}\int k_B TdN.
\end{equation}
In the above equation $T$ is the temperature of the holographic
screen. Following Unruh's argument, as well as the relation
between $\phi$ and $a$, we get \cite{Ver}
\begin{eqnarray} \label{KbT}
k_BT=\frac{\hbar}{2\pi c}e^{\phi}N^b\nabla_b\phi,
\end{eqnarray}
where $N^b$ is an outward pointing vector which is normal to both
the screen $\Sigma_n$ and $\xi^{\alpha}$ \cite{Ver}. Inserting
(\ref{KbT}) into Eq. (\ref{mass}), we find
\begin{eqnarray}
M=\frac{\hbar}{2\pi c^3}\int e^{\phi}N^b\nabla_b\phi dN.
\end{eqnarray}
Using the relation between $\phi$ and $\xi^{\alpha}$, as well as
Stokes theorem, we find \cite{Ver,wald}
\begin{eqnarray}
M=\frac{\hbar}{4\pi c^3 \ell_p^{n-1}}\int R_{\mu \nu}n^{\mu}\xi^{\nu}dV_n,
\end{eqnarray}
where $n^{\mu}$ is a normal to the volume $V_n$. Combining this
result with Eq. (\ref{G}) yields
\begin{eqnarray}\label{mass1}
M=\frac{\Gamma(\frac{n}{2})}{2 \pi^{\frac{n}{2}}G_{n+1}}\int
R_{\mu \nu}n^{\mu}\xi^{\nu}dV_n.
\end{eqnarray}
Therefore, the $M\geq0$ condition is similar to the Averaged Strong
energy condition (ASEC) \cite{energy}. For $n=3$ this relation is
reduced to
\begin{eqnarray}\label{em}
M=\frac{1}{4 \pi G}\int R_{\mu \nu}n^{\mu}\xi^{\nu}dV_3,
\end{eqnarray}
where $G \equiv G_4={c^3\ell_p^2}/{\hbar}$,  is the
four-dimensional Newtonian gravitational constant and so, previous
result is recovered \cite{Ver,wald}.

Nowadays, it is generally accepted that the active gravitational
mass (Komar mass) plays the role of the mass \cite{Pad3}. The
Komar mass in $(n+1)$-dimensions is defined as \cite{Cai4,Cai41}
\begin{equation}\label{M1}
\mathcal M =\frac{n-1}{n-2}
\int_{V_n}{dV_n\left(T_{\mu\nu}-\frac{1}{n-1}Tg_{\mu\nu}\right)n^{\mu}\xi^{\nu}}.
\end{equation}
It is useful to mention that if the right hand side of this equation
is positive, then we face with positive energies ($\mathcal M\geq0$)
which is also similar to ASEC \cite{energy}. In this equation,
$T_{\mu \nu}$ is the energy momentum tensor in ($n+1$)-dimension. By
equating (\ref{mass1}) and (\ref{M1}) we get
\begin{equation}\label{eins}
R_{\mu
\nu}=\frac{2(n-1)\pi^{\frac{n}{2}}G_{n+1}}{(n-2)\Gamma(\frac{n}{2})}
(T_{\mu\nu}-\frac{1}{n-1}Tg_{\mu\nu}).
\end{equation}
So, the ($3+1$)-dimensional Einstein equation will be restored by
choosing $n=3$. The relation between Einstein gravitational
constant in $(n+1)-$dimensions, $\kappa_{n+1}$, and the Newtonian
constant $G_{n+1}$ can be written as \cite{mansori,SMR}
\begin{equation}\label{kn}
\kappa_{n+1}=\frac{2(n-1)\pi^{n/2}
G_{n+1}}{(n-2)(\frac{n}{2}-1)!}.
\end{equation}
Substituting (\ref{kn}) in (\ref{eins}), we obtain
\begin{eqnarray}
R_{\mu
\nu}=\kappa_{n+1}\left(T_{\mu\nu}-\frac{1}{n-1}Tg_{\mu\nu}\right),
\end{eqnarray}
which can also be rewritten as
\begin{equation}\label{EQS}
R_{\mu \nu}-\frac{1}{2}Rg_{\mu \nu}=\kappa_{n+1} T_{\mu \nu}.
\end{equation}
In this equation, $R_{\mu \nu}$ and $T_{\mu \nu}$ are Ricci tensor
and energy momentum in ($n+1$)-dimensions, respectively. For $n=3$
we have $\kappa_4=8\pi G$ and the well-known coefficient of the
Einstein equation is restored \cite{gron}. Therefore, Eq. (\ref{EQS}) is
nothing but the ($n+1$)-dimensional Einstein equations. Now,
consider an ($4+1$)-dimensional manifold which is described by the
line element
\begin{eqnarray}\label{line} ds^2=d\zeta^2+q_{\mu
\nu}dx^{\mu}dx^{\nu}.
\end{eqnarray}
In the above equation, $q_{\mu \nu}$ is the metric of the
($3+1$)-dimensional submanifold $\zeta=\rm constant$. Without lose
of generality, we can choose the hypersurface $\zeta=0$. Hence, we
have
\begin{eqnarray}
g_{\mu \nu}=q_{\mu \nu}+n_{\mu}n_{\nu},
\end{eqnarray}
where $n^{\mu}=\delta^{\mu}_{\zeta}$ is the spacelike unit normal
vector to the ($3+1$)-dimensional submanifold. We assume the
energy momentum tensor of the ($4+1$)-dimensional bulk  has the
following form \cite{shim}
\begin{eqnarray}\label{EMT}
^5T_{\mu \nu}=-\Lambda g_{\mu \nu}+\delta(\zeta)S_{\mu \nu}.
\end{eqnarray}
In this equation, $\Lambda=-\frac{6}{\ell^2}$ is the bulk
cosmological constant where $\ell$ is the curvature radii of the
bulk. In addition, since for $r<\ell$ the effects of the extra
dimension may have an acceptive contribution to the gravity compared
with those of the ordinary dimensions, $\ell$ can be also considered
as the effective size of the extra dimension \cite{ellis}. In
addition, $S_{\mu \nu}$ is decomposed into two parts including the
energy momentum tensor of the ordinary matter ($\tau_{\mu \nu}$) and
the tension ($\lambda$) of the $3$-brane,
\begin{eqnarray}\label{brane}
S_{\mu \nu}=\lambda q_{\mu \nu} +\tau_{\mu \nu}.
\end{eqnarray}
From this expression for the energy momentum tensor of the bulk, it
is apparent that, unlike the other matter field, gravity can
penetrate into the fifth dimension. We should note that $M_p$ is the
Planck mass and $\lambda=\frac{3M_p^2}{4\pi\ell^2}$. In addition, it
seems that the fifth dimension may affect the Newton's law of
gravity when the effective size of the extra dimension satisfies the
$\ell\leq0.1$~mm condition, and therefore one gets the
$\Lambda\leq-6\times10^{8}$eV and $\lambda>(1~\textrm{TeV})^4$
limits for the bulk cosmological constant and the brane tension
respectively \cite{ellis,Ul}. This setup, which was originally
proposed by Randall and Sundrum \cite{RS}, can explain the weakness
of the gravity against the other fundamental forces \cite{RS}. The
key point in obtaining the Einstein equations on the brane is
reflecting the ($4+1$)-dimensional Einstein equations on the
$3$-brane, which is a mathematical problem \cite{shim}. In fact, the
Einstein equations for the bulk can be written as
\begin{equation}\label{EQS1}
^5G_{\mu \nu}\equiv \ ^5R_{\mu \nu}-\frac{1}{2}\ ^5Rg_{\mu
\nu}=\kappa_5 \ ^5T_{\mu \nu}.
\end{equation}
Also from (\ref{EMT}), the Israel's junction conditions read \cite{shim}
\begin{eqnarray}\label{jc1}
[q_{\mu \nu}]=0\ , \    \    \   [K_{\mu
\nu}]=-\kappa_5\left(S_{\mu \nu}-\frac{S}{3}q_{\mu \nu}\right),
\end{eqnarray}
where $[A]\equiv \lim_{\zeta\rightarrow
C^+}A-\lim_{\zeta\rightarrow C^-}A$. In order to find
the projection of (\ref{EQS1}) on the brane, we multiply
(\ref{EQS1}) by $q^{\mu}_{\beta}q^{\nu}_{\delta}$,
\begin{equation}\label{EQS2}
^5G_{\mu \nu}q^{\mu}_{\beta}q^{\nu}_{\delta}=\kappa_5 \ ^5T_{\mu
\nu}q^{\mu}_{\beta}q^{\nu}_{\delta}.
\end{equation}
Taking into account the $Z_2$ symmetry of the bulk, and using the
relations between Einstein tensor in four and five dimensions as
well as the Israel's junction conditions, and following the
approach of the authors in \cite{shim}, we arrive at
\begin{eqnarray}\label{Einbrane}
^4G_{\beta \delta}=8\pi G_N\ ^4T_{\beta \delta}.
\end{eqnarray}
In deriving above equation, we have used the following definitions
\begin{eqnarray}\label{T}
\ ^4T_{\beta \delta}&\equiv&\tau^{\prime}_{\beta
\delta}+\frac{6}{\lambda}\Pi_{\beta \delta}-\frac{6}{\lambda
\kappa_5^2} E_{\beta \delta},\\ \nonumber \tau^{\prime}_{\beta
\delta}&\equiv&\tau_{\beta \delta}-Qq_{\beta \delta},
\end{eqnarray}
where
\begin{eqnarray}\label{TTU}
\Pi_{\beta \delta}&\equiv&-\frac{1}{4}\tau_{\beta \alpha}\tau
^{\alpha}_{\delta}+\frac{1}{12}\tau \tau_{\beta
\delta}+\frac{1}{8}q_{\beta \delta}\tau_{\mu \nu}\tau^{\mu
\nu}-\frac{1}{24} q_{\beta \delta} \tau^2,
\end{eqnarray}
and
\begin{eqnarray}\label{cons}
Q=\frac{3\Lambda}{\kappa_5 \lambda}+\frac{\lambda}{2}, \  \  \
G_N=\frac{\kappa_5^2 \lambda}{48\pi}.
\end{eqnarray}
Indeed, $\Pi_{\mu\nu}$ is a geometrical term coming from the
extrinsic curvature terms, which is written in the~(\ref{TTU}) form,
by considering the Israel junction conditions along as the
($4+1$)-dimensional Einstein equations \cite{shim,mead}. Also
$E_{\beta \delta}=\ ^5C^{\alpha}_{\mu \nu
\gamma}n_{\alpha}n^{\nu}q^{\mu}_{\beta}q^{\gamma}_{\delta}$ is a
traceless tensor including the projection of the five dimensional
Weyl tensor ($\ ^5C^{\alpha}_{\mu \nu \gamma}$) onto the brane
meaning that $E_{\beta \delta}n^{\beta}=0$ \cite{marteen}. In
addition, $Q$ and $G_N$ are the vacuum energy density and the
Newton's gravitational constant on the brane respectively.
Therefore, in this way we obtain the Einstein equations onto the
brane by reflecting the ($4+1$)-dimensional Einstein field equations
on the ($3+1$)-dimensional hypersurface. In addition, $E_{\beta
\delta}$ is the limiting value at either $\zeta\rightarrow 0^{+}$ or
$\zeta\rightarrow 0^{-}$. It is worth to mention that $G_N$ differs
from the ordinary Newtonian gravitational constant in
($3+1$)-dimensional spacetime (\ref{G}). Also, the relation between
$G_N$ and $\kappa_5$ differs from (\ref{kn}). In addition, for
$\lambda\leq0$, we see that the Newtonian gravitational constant on
the brane is misdefined. In fact, these differences originates from
the projective nature of the Einstein equations onto the brane.
Finally, we should note that Eq.~(\ref{Einbrane}) converges to the
general relativity, provided we neglect the bulk effects, namely we
take the $\kappa_5\rightarrow0$ limit while keeping $G_N$ finite
\cite{shim}.

Bianchi identity implies $D_{\beta}~^4G^{\beta\delta}=0$, where
$D_{\beta}$ denotes the covariant derivative in $4$-dimensional
spacetime, which leads to
\begin{eqnarray}\label{ec1}
D_{\beta}~^4T^{\beta \delta}=0.
\end{eqnarray}
Whenever the ordinary matter fields ($\tau_{\beta\delta}$) are only
distributed on the brane~(\ref{EMT}), since $D_{\mu}g^{\mu\nu}=0$,
we get $D_{\beta}\tau^{\beta\delta}=0$ and
\begin{eqnarray}\label{ec2}
D_{\beta}E^{\beta \delta}=\kappa_5^2D_{\beta}\Pi^{\beta \delta}.
\end{eqnarray}
Indeed, Eqs.~(\ref{ec1}) and~(\ref{ec2}) are nothing but the
conservation equations on the brane \cite{shim,meadea}. It is
useful to note that for a pure anti de-Sitter bulk we reach
\begin{eqnarray}\label{ec3}
D_{\beta}\Pi^{\beta \delta}=0,
\end{eqnarray}
since $E_{\beta\delta}=0$ \cite{meadea}. The energy-momentum
tensor of a prefect fluid source is seen by an observer with four
velocity $u^{\nu}$ as
\begin{eqnarray}\label{pf}
\tau_{\mu}^{\nu}=(\rho+p)u_{\mu}u^{\nu}+p\delta_{\mu}^{\nu},
\end{eqnarray}
leading to~(\ref{TTU})
\begin{eqnarray}\label{pf1}
\Pi_{\mu}^{\nu}=\frac{\rho}{6}[(\rho+p)u_{\mu}u^{\nu}+(\frac{\rho}{2}+p)\delta_{\mu}^{\nu}],
\end{eqnarray}
where we have considered the ($-~+~+~+$) signature for the brane
metric. Using Eq.~(\ref{ec3}), it is easy to show that an
inhomogeneous prefect fluid is rejected whenever, the bulk is
purely anti de-Sitter \cite{marteen,meadea}. For any non-spacelike
observer, which moves onto the brane, with four velocity
$u^{\nu}$, the energy density should be positive which is called
the weak energy condition (WEC) \cite{pois,energy}. Since
$\tau_{\mu\nu}$ carries the information of energy source, WEC
implies that $\tau_{\mu\nu}u^{\mu}u^{\nu}\geq0$ leading to
\begin{eqnarray}\label{wec}
\rho\geq0,~~\rho+p>0,
\end{eqnarray}
where we have used~(\ref{pf}) to get this relation
\cite{hawkellis,pois}. Applying WEC to $\Pi_{\mu}^{\nu}$
introduced in~(\ref{pf1}), we get
\begin{eqnarray}
\frac{\rho^2}{12}\geq0,~~\frac{\rho}{6}(\rho+p)>0.
\end{eqnarray}
It is apparent that if WEC is satisfied by $\tau_{\mu\nu}$, then
$\Pi_{\mu\nu}$ will also satisfy WEC. It is straightforward to see
that the $-Qg_{\mu\nu}$ term satisfies $\rho\geq0$ for $Q>0$, and
does not satisfy $\rho+p>0$ ($\rho+p=Q-Q=0$). Therefore, WEC is not
fully satisfied by $-Qg_{\mu\nu}$. Additionally, WEC is completely
violated by $-Qg_{\mu\nu}$ whiles $Q<0$. From geometrical point of
view, WEC implies that $~^4G_{\mu \nu}u^{\mu}u^{\nu}\geq0$ which
leads to the $~^4T_{\mu \nu}u^{\mu}u^{\nu}\geq0$ condition
\cite{energy}. It is also useful to mention that WEC may be violated
by the $E_{\mu\nu}$ term \cite{neg}. The latter my lead to
$~^4T_{\mu \nu}u^{\mu}u^{\nu}<0$ telling us that WEC can be violated
by the $~^4T_{\mu \nu}$ term, independent of $-Qg_{\mu\nu}$
\cite{neg}. Moreover, since a null observer is a non-spacelike
observer, by continuity we can conclude that the energy density
corresponding to a null observer, with tangent vector field
$k^{\nu}$, should be positive meaning that
$\tau_{\mu\nu}k^{\mu}k^{\nu}\geq0$
\cite{wald,hawkellis,pois,energy}.
$~^4G_{\mu\nu}k^{\mu}k^{\nu}\geq0$ is the geometrical interpretation
of this energy condition leading to
$~^4T_{\mu\nu}k^{\mu}k^{\nu}\geq0$ \cite{energy}. By applying this
condition to~(\ref{pf}), we get
\begin{eqnarray}\label{wec}
\rho\geq0,~~\rho+p\geq0.
\end{eqnarray}
Indeed, WEC implies NEC \cite{wald,hawkellis,pois,energy}. For the
$\Pi_{\mu\nu}$ term, we reach
\begin{eqnarray}
\frac{\rho^2}{12}\geq0,~~\frac{\rho}{6}(\rho+p)\geq0,
\end{eqnarray}
where we have used~(\ref{pf1}) to obtain this equation. Therefore,
if NEC is satisfied by $\tau_{\mu\nu}$, then NEC is also satisfied
by $\Pi_{\mu\nu}$. Moreover, NEC is marginally satisfied by the
$-Qg_{\mu\nu}$ term for $Q>0$, and is not satisfied for $Q<0$. As
the WEC case, NEC may be violated by the $E_{\mu\nu}$ term and thus
the $~^4T_{\mu \nu}$ term \cite{neg}.

Physically, Dominant Energy Condition (DEC) implies that the density
of matter momentum ($-\tau^{\mu\nu}u_{\mu}$) measured by an observer
with a four velocity $u^{\mu}$ should be non-spacelike
\cite{pois,hawkellis}. Applying this condition to~(\ref{pf}), one
gets $\rho\geq0$ and $|p|\leq\rho$ \cite{pois}. Calculations
for~(\ref{pf1}) leads to $\rho\geq0$ and $|1+\frac{2p}{\rho}|\leq1$.
The latter means that $p$ should either satisfy the $p\leq0$ or
$-p\leq\rho$ conditions. Therefore, for $0\leq p\leq\rho$,
$\tau_{\mu\nu}$ and $\Pi_{\mu\nu}$ satisfy WEC, NEC and DEC
simultaneously. Moreover, it is straightforward to see that the
$-Qg_{\mu\nu}$ term will marginally satisfy DEC, only for $Q>0$.
From geometrical point of view, DEC implies that
$-~^4G^{\mu\nu}u_{\mu}$ should be causal \cite{energy}. By using
this geometrical interpretation, it is shown that $E_{\mu\nu}$ may
violate DEC which can lead to violate DEC by $~^4G^{\mu\nu}$ and
thus $~^4T_{\mu \nu}$ \cite{neg}.

In order to derive a suitable criterion for studying the Strong
Energy Condition (SEC), we should write the Raychaudhuri equation
for a congruence of timelike geodesics on the brane as
\cite{marteen}
\begin{eqnarray}
\frac{d\Theta}{d\lambda}+\frac{1}{3}\Theta^2+\sigma_{\mu\nu}\sigma^{\mu\nu}-\omega_{\mu\nu}\omega^{\mu\nu}
+8\pi G_N(\tau_{\mu\nu}-\frac{1}{2}\tau
q_{\mu\nu})u^{\mu}u^{\nu}-8\pi
G_NQ=-\frac{1}{12}\kappa_5^2(\rho(2\rho+3p)).
\end{eqnarray}
In deriving the above equation, we have assumed that
$\tau_{\mu\nu}$ has the prefect fluid form, and did the
calculations for a purely anti de-Sitter bulk meaning that
$E_{\mu\nu}=0$. These considerations lead to a homogeneous density
and thus homogeneous pressure which implies that the four
acceleration vector of congruence of the timelike geodesics is
zero \cite{mead,ellis1}. Now, using Eq.~(\ref{cons}) to get:
\begin{eqnarray}\label{r1}
\frac{d\Theta}{d\lambda}+\frac{1}{3}\Theta^2+\sigma_{\mu\nu}\sigma^{\mu\nu}-\omega_{\mu\nu}\omega^{\mu\nu}
+8\pi G_N(\tau^{\prime}_{\mu\nu}-\frac{1}{2}\tau^{\prime} q_{\mu\nu})u^{\mu}u^{\nu}=-\frac{1}{12}\kappa_5^2(\rho(2\rho+3p)).
\end{eqnarray}
It is also easy to show that the right hand side of this equation can be written as:
\begin{eqnarray}\label{r2}
\frac{1}{12}(\rho(2\rho+3p))=(\Pi_{\mu\nu}-\frac{1}{2}\Pi q_{\mu\nu})u^{\mu}u^{\nu}.
\end{eqnarray}
By combining Eqs.~(\ref{r1}) and~(\ref{r2}), and using the Einstein equations onto the brane~(\ref{Einbrane}) we get
\begin{eqnarray}\label{r3}
\frac{d\Theta}{d\lambda}=-\frac{1}{3}\Theta^2-\sigma_{\mu\nu}\sigma^{\mu\nu}+\omega_{\mu\nu}\omega^{\mu\nu}
-~^4R_{\mu\nu}u^{\mu}u^{\nu}.
\end{eqnarray}
It is apparent that when $\kappa_5\rightarrow0$, while
$G_N\rightarrow c\neq0$, this equation converges to that of the
Einstein theory \cite{energy,marteen}). Since for the hypersurface
orthogonal congruences $\omega_{\mu\nu}=0$, just the same as the GR,
the attractive nature of the gravity implies
$~^4R_{\mu\nu}u^{\mu}u^{\nu}\geq0$ \cite{mead}. In fact, it is shown
that for a general situation, including $E_{\mu\nu}\neq0$ and
arbitrary source of energy, $~^4R_{\mu\nu}u^{\mu}u^{\nu}\geq0$
leading to $(~^4T_{\mu\nu}-\frac{1}{2}~^4T
q_{\mu\nu})u^{\mu}u^{\nu}\geq0$ can be considered as a suitable
criterion for studying SEC \cite{mead}. It is also useful to mention
here that the $E_{\mu\nu}$ term may lead to violate SEC \cite{neg}.
More comprehensive notes about the energy conditions as well as the
matter evolution in this theory can be found in
\cite{mead,marteen,neg}.

In order to calculate the induced Komar mass on the brane, bearing
the traceless nature of $E_{\beta \delta}$ in mind. In addition,
we insert $n=3$ into Eq.~(\ref{M1}), after using Eq.~(\ref{T}), we reach at
\begin{eqnarray}\label{M2}
\mathcal M=2\int_{V_3}dV_3\left[\left(\tau^{\prime}_{\mu
\nu}+\frac{6}{\lambda}\Pi_{\mu \nu}-\frac{6}{\lambda
\kappa_5^2} E_{\mu
\nu}\right)-\frac{1}{2}\left(\tau^{\prime}+\frac{6}{\lambda}\Pi\right)q_{\mu
\nu}\right]n^{\mu}\xi^{\nu}.
\end{eqnarray}
In this equation, $n^{\mu}$ and $\xi^{\nu}$ are the normal unit
vector to the volume $V_3$ and the timelike Killing vector on the
brane, respectively. It is useful to mention that for the positive
values of RHS of this equation we get $\mathcal M\geq0$ which is
similar to ASEC \cite{energy}. From Eq.~(\ref{line}) we see that
$E_{\mu \nu}$ vanishes either for the flat brane spacetime ($q_{\mu
\nu}$) or the Ads bulk \cite{shim,mead}. This leads us to
\begin{eqnarray}\label{M2n}
\mathcal M=2\int_{V_3}dV_3\left[\left(\tau^{\prime}_{\mu
\nu}+\frac{6}{\lambda}\Pi_{\mu
\nu}\right)-\frac{1}{2}\left(\tau^{\prime}+\frac{6}{\lambda}\Pi\right)q_{\mu
\nu}\right]n^{\mu}\xi^{\nu}.
\end{eqnarray}
Such condition $(E_{\mu \nu}=0)$ leads to
$\partial_{\mu}T^{\mu\nu}=0$, implying that the energy-momentum
conservation law on the brane is the same as that of the Einstein
gravity only for the AdS$_5$ bulk \cite{shim}. Assuming the
Randall-Sundrum fine-tuning, $Q=\frac{3\Lambda}{\kappa_5
\lambda}+\frac{\lambda}{2}=0$, holds on the brane one can easily
check that
\begin{eqnarray}\label{M2n}
\mathcal M=2\int_{V_3}dV_3\left[\left(\tau_{\mu
\nu}+\frac{6}{\lambda}\Pi_{\mu
\nu}\right)-\frac{1}{2}\left(\tau+\frac{6}{\lambda}\Pi\right)q_{\mu
\nu}\right]n^{\mu}\xi^{\nu},
\end{eqnarray}
which is compatible with the results obtained in
\cite{Sheykhi24,Ling}. Therefore, unlike Eq.~(\ref{M2}), the Komar
mass definition used in \cite{Sheykhi24,Ling} is confined to the
brane satisfying the $E_{\mu \nu}=0$ and $Q=0$ conditions
simultaneously. Finally, we should note that our relation for the
Komar mass (\ref{M2}) is more comprehensive than Eq.~(\ref{M2n})
introduced in \cite{Sheykhi24,Ling}. Equating Komar definition for
the mass in (\ref{M2n}) with Eq.~(\ref{em}) we finally obtain
\begin{eqnarray}\label{eeb}
G_{\mu \nu}= 8 \pi G\left(\tau_{\mu \nu}+\frac{6}{\lambda}\Pi_{\mu
\nu}\right).
\end{eqnarray}
Bearing $E_{\mu \nu}=0$ and $Q=0$ in mind, we see that this equation
is similar to the Einstein field equations on the
brane~(\ref{Einbrane}), provided we take $G=G_N$ yielding
$\Lambda=-\frac{4}{3\ell_p}$. The negative sign in
$\Lambda=-\frac{4}{3\ell_p}$ is signalling that the bulk spacetime
should be AdS. Indeed, comparing this equation with
Eq.~(\ref{Einbrane}), we find that Eq.~(\ref{eeb}) is valid if the
bulk spacetime is AdS. Finally, We should also note that the Komar
mass definition~(\ref{M2n}), introduced in \cite{Sheykhi24,Ling},
along as the entropic force scenario cannot lead to the true
Einstein field equations on the brane.
\section{conclusions and discussions}
We have considered the higher dimensional definition for the Komar
mass, and by taking into account the entropic origin for gravity, we
derive the Einstein field equations in arbitrary dimensions. This
procedure naturally leads to the derivation of the higher
dimensional gravitational coupling constant of the Einstein
equations which is in complete agreement with the results obtained
by comparing the weak field limit of the Einstein equations with
Poisson equation in higher dimensions \cite{SMR,mansori}. We
mentioned that one can find the Einstein equations onto the brane by
reflecting the Einstein equations on the ($3+1$)-dimensional
submanifold. The quality of availability of some energy conditions,
including WEC, NEC, DEC and SEC, were briefly studied. Then, we
discussed the differences between Newtonian gravitational constant
on the brane, $G_N$, and the ordinary Newtonian gravitational
constant ($G$). This difference originates from the higher
dimensional considerations and projective nature of the Einstein
equation on the brane. In addition, we have derived an expression
for the Komar mass on the brane, and compared our relation with that
of used in previous works \cite{Ling,Sheykhi24}. Finally, by
considering the entropic origin for the gravity and employing the
Komar definition for mass \cite{Ling,Sheykhi24}, we found the
Einstein field equations onto the $3$-brane embedded in a five
dimensional AdS bulk. In addition, we pointed to the weaknesses of
the Einstein field equations on the brane when the Komar mass
definition introduced in \cite{Ling,Sheykhi24} is used.
\acknowledgments{We are grateful to the respected referee for
valuable comments leading to improve this manuscript. A. Sheykhi
thanks Shiraz University Research Council. This work has been
supported financially by Research Institute for Astronomy \&
Astrophysics of Maragha (RIAAM), Iran.


\begin{thebibliography}{99}
\bibitem{B1} J. D. Bekenstein, Phys. Rev. D {\bf7}, 2333 (1973).
\bibitem{H1} S. W. Hawking., Commun Math. Phys. {\bf43}, 199 (1975).
\bibitem{H2} S. W. Hawking., Nature {\bf248}, 30 (1974).
\bibitem{B2} J. M. Bardeen, B. Carter, S. W. Hawking, Commun. Math. Phys. {\bf31}, 161 (1973).
\bibitem{D1} P. C. W. Davies, J. Phys. A: Math. Gen. {\bf8}, 609 (1975).
\bibitem{u1} W. G .Unruh, Phys. Rev. D {\bf14}, 870 (1976).
\bibitem{s1} L. Susskind, J. Math. Phys. {\bf36}, 6377 (1995).
\bibitem{Jac} T. Jacobson, Phys. Rev. Lett. {\bf75}, 1260 (1995).
\bibitem{Elin} C. Eling, R. Guedens, T. Jacobson, Phys. Rev. Lett. {\bf96}, 121301 (2006).
\bibitem{Pad} T. Padmanabhan, Class. Quantum. Grav. {\bf19}, 5387 (2002).
\bibitem[\protect\citeauthoryear{M.~Akbar et al.}{2007}]{Cai2} M. Akbar, R. G. Cai, Phys. Rev. D {\bf 75}, 084003 (2007).
\bibitem[\protect\citeauthoryear{R.~G.~Cai et al.}{2007}]{Cai3} R. G. Cai, L. M. Cao, Phys.Rev. D {\bf 75}, 064008 (2007).
\bibitem[\protect\citeauthoryear{R.~G.~Cai et al.}{2005}]{CaiKim} R. G. Cai, S. P. Kim, JHEP {\bf0502}, 050
(2005).
\bibitem[\protect\citeauthoryear{B. Wang et al.}{2001}]{W1} B. Wang, E. Abdalla, K. R. Su, Phys.Lett. B {\bf503},  394 (2001).
\bibitem[\protect\citeauthoryear{B. Wang et al.}{2002}]{W2} B. Wang, E. Abdalla, K. R. Su, Mod. Phys. Lett. A {\bf17},  23 (2002).
\bibitem[\protect\citeauthoryear{R.~G.~Cai et al}{2008}]{Cai33} R. G. Cai, M. L. Cao, P. Y. Hu,  JHEP {\bf0808}, 090, (2008).
\bibitem[\protect\citeauthoryear{S. Nojiri et al}{2006}]{Shey0} S. Nojiri, S. D. Odintsov, Gen. Relativ. Gravit. {\bf38}, 1285 (2006).
\bibitem[\protect\citeauthoryear{A .Sheykhi}{2010}]{Shey01} A. Sheykhi, Class. Quantum. Grav. {\bf27}, 025007 (2010).
\bibitem[\protect\citeauthoryear{A. Sheykhi.}{2010}]{Shey02} A. Sheykhi, Eur. Phys. J. C {\bf69}, 265 (2010).
\bibitem[\protect\citeauthoryear{A. Sheykhi}{2007}]{Shey1} A. Sheykhi, B. Wang, R. G. Cai, Nucl. Phys. B {\bf
779}, 1 (2007).
\bibitem[\protect\citeauthoryear{R.~G.~Cai. et. al}{2007}]{C1} R. G. Cai, L. M. Cao, Nucl. Phys. B {\bf785}, 135 (2007).
\bibitem[\protect\citeauthoryear{A. Sheykhi et. al}{2007}]{Shey2} A. Sheykhi, B. Wang, R. G. Cai, Phys. Rev. D {\bf76}, 023515 (2007).
\bibitem[\protect\citeauthoryear{A. Sheykhi. et al}{2009}]{Shey22} A. Sheykhi, B. Wang, Phys. Lett. B {\bf678}, 434 (2009).
\bibitem[\protect\citeauthoryear{A. Sheykhi}{2009}]{Shey23} A. Sheykhi, JCAP {\bf05}, 019 (2009).
\bibitem[\protect\citeauthoryear{T. Padmanabhan}{2010}]{Pad0} T. Padmanabhan, Rept. Prog. Phys. {\bf73}, 046901 (2010).
\bibitem{Padm} T. Padmanabhan, Mod. Phys. Lett. A {\bf25}, 1129 (2010).
\bibitem{Ver} E. Verlinde, JHEP {\bf1104}, 029 (2011).
\bibitem[\protect\citeauthoryear{R.G. Cai. et. al}{2010}]{Cai4} R. G. Cai, L. M. Cao, N. Ohta, Phys. Rev. D {\bf81}, 061501(R) (2010).
\bibitem[\protect\citeauthoryear{R.G. Cai et. al}{2010}]{Cai41} R. G. Cai, L. M. Cao, N. Ohta, Phys. Rev. D {\bf81}, 084012 (2010).
\bibitem[\protect\citeauthoryear{L. Smolin}{2010}]{Smolin} L. Smolin, arXiv:1001.3668.
\bibitem[\protect\citeauthoryear{M. Li. et. al}{2010}]{Li} M. Li, Y. Wang, Phys. Lett. B {\bf687}, 243 (2010).
\bibitem[\protect\citeauthoryear{Y. Tian et. al}{2010}]{Tian} Y. Tian, X. Wu, Phys. Rev. D \textbf{81}, 104013 (2010).
\bibitem[\protect\citeauthoryear{Y. S. Myung}{2010}]{Myung1} Y. S. Myung, arXiv:1002.0871.
\bibitem[\protect\citeauthoryear{I. V. Vancea. et. al}{2010}]{Vancea} I. V. Vancea, M. A. Santos., arXiv:1002.2454.
\bibitem[\protect\citeauthoryear{L. Modesto. et. al}{2010}]{Modesto} L. Modesto, A. Randono., arXiv:1003.1998.
\bibitem[\protect\citeauthoryear{A. Sheykhi}{2010}]{Sheykhi1} A. Sheykhi, Phys. Rev. D \textbf{81}, 104011 (2010).
\bibitem[\protect\citeauthoryear{B. Liu. et. al}{2011}]{BLi} B. Liu, Y. C. Dai, X. R. Hu, J. B. Deng, Mod. Phys. Lett. A {\bf26}, 489 (2011).
\bibitem[\protect\citeauthoryear{S. H. Hendi et. al}{2011}]{Sheykhi2} S. H. Hendi, A. Sheykhi, Phys.  Rev.  D {\bf83}, 084012 (2011).
\bibitem[\protect\citeauthoryear{A. Sheykhi et. al}{2011}]{Sheykhi21} A. Sheykhi, S. H. Hendi, Phys.  Rev.  D {\bf84}, 044023 (2011).
\bibitem[\protect\citeauthoryear{S. H. Hendi et. al}{2012}]{Sheykhi22} S. H. Hendi, A. Sheykhi, Int. J Theor. Phys. {\bf51}, 1125 (2012).
\bibitem[\protect\citeauthoryear{A. Sheykhi et. al}{2012}]{Sheykhi23} A. Sheykhi, Z. Teimoori, Gen. Relativ. Gravit. {\bf44}, 1129 (2012).
\bibitem{Ling} Y. Ling, J. P. Wu, J. Cosmol. Astropart. Phys. {\bf1008}, 017 (2010).
\bibitem[\protect\citeauthoryear{A. Sheykhi}{2012}]{Sheykhi24} A. Sheykhi, Int. J Theor. Phys. {\bf51}, 185 (2012).
\bibitem[\protect\citeauthoryear{W. Gu. et. al}{2010}]{Gu} W. Gu, M. Li, R. X. Miao, arXiv:1011.3419.
\bibitem[\protect\citeauthoryear{R. X. Miao et. al}{2011}]{Miao1} R. X. Miao, J. Meng, M. Li, arXiv:1102.1166.
\bibitem[\protect\citeauthoryear{Y. X. Liu et. al}{2010}]{other} Y. X. Liu, Y. Q. Wang, S. W. Wei, Class. Quantum
Grav. \textbf{27}, 185002 (2010).
\bibitem[\protect\citeauthoryear{R. B. Mann}{2011}]{mann} R. B. Mann, J. R. Mureika, Phys. Lett. B {\bf703}, 167 (2011).
\bibitem{SMR} A. Sheykhi, H. Moradpour, N. Riazi, Gen. Rel. Grav. {\bf45}, 1033 (2013).
\bibitem{Pol} J. Polchinski, \textit{String Theory I \& II} (Cambridge Univ. Press, Cambridge, 1998).
\bibitem{Mad} J. M. Maldacena, Adv. Theor. Math. Phys. {\bf2}, 231 (1998).
\bibitem{Mad1} J. M. Maldacena, Int. J. Theor. Phys. {\bf38}, 1113 (1999).
\bibitem{Sus} L. Susskind, J. Math. Phys. {\bf36}, 6377 (1995).
\bibitem{Bous} R. Bousso, Rev. Mod. Phys. {\bf74}, 825 (2002).
\bibitem{RS1} L. Randall, R. Sundrum, Phys. Rev. Lett. {\bf83}, 3370 (1999).
\bibitem{RS} L. Randall, R. Sundrum, Phys. Rev. Lett. {\bf83}, 4690 (1999).
\bibitem{shim} T. Shiromizu, K. Maeda and M. Sasaki, Phys. Rev. D \textbf{62}, 024012 (2000).
\bibitem{ppp} A. Paranjape, S. Sarkar and T. Padmanabhan, Phys. Rev. D \textbf{74}, 104015 (2006).
\bibitem{Pad1} T. Padmanabhan, Phys. Rev. D {\bf81}, 124040 (2010).
\bibitem{wald} R. M. Wald, \textit{General Relativity} (The University of Chicago Press, 1984).
\bibitem{energy} E. Curiel,  arXiv:1405.0403.
\bibitem{Pad3} T. Padmanabhan, Class. Quantum. Grav. {\bf21}, 4485 (2004).
\bibitem{mansori} R. Mansouri, A. Nayeri, Gravitation Cosmology, {\bf4}, No. 2, 142 (1998).
\bibitem{gron} O. Gron and S. Hervik, \textit{Einstein's General Theory of Relativity} (Springer LLC, 2007).
\bibitem{ellis} G. F. R. Ellis, R. Marteens and M. A. H. Maccallum, \textit{Relativistic Cosmology}
(Cambridge Univ. Press, New York, 2012).
\bibitem{Ul} U. Ellwanger, Mod. Phys. Lett. A {\bf 20}, 2521 (2005).
\bibitem{mead} R. Marteens, Living Rev. Rel. {\bf 7}, 7 (2004).
\bibitem{marteen} R. Marteens, Phys. Rev. D {\bf62}, 084023 (2000).
\bibitem{meadea} M. Sasaki, T. Shiromizu, and K. Maeda, Phys. Rev. D {\bf62}, 024008 (2000).
\bibitem{pois} E. Poisson, \textit{A Relativist's Toolkit} (Cambridge University Press, UK, 2004).
\bibitem{hawkellis} S. W. Hawking and G. F. R. Ellis, \textit{the large scale structure of space-time},
(Cambridge University Press, USA, 1973).
\bibitem{neg} D. N. Vollick, Gen. Rel. Grav. {\bf 34}, 1 (2002).
\bibitem{ellis1} G. F. R. Ellis, and H. van Elst, NATOAdv. StudyInst. Ser. C. Math. Phys. Sci. {\bf 541}, 1 (1999) [arXiv:gr-qc/9812046].
\end{thebibliography}
\end{document}